\documentclass[prd,aps,preprint,tightenlines,nofootinbib,superscriptaddress]{revtex4}
\usepackage{epsfig}
\usepackage{amsmath, amssymb, amsfonts}
\newcommand{\p}{\partial}

\newcommand{\ma}{\mathbf}

\newcommand{\boldgreek}[1]{{\mbox{\boldmath{$#1$}}}}
\newcommand{\bg}{\boldgreek}

\begin{document}

\title{Angular Momentum in Non-Relativistic QED \\ and Photon Contribution to Spin of Hydrogen Atom}

\author{Panying Chen}
\affiliation{Maryland Center for Fundamental Physics, Department of Physics, \\ University of Maryland, College
Park, Maryland 20742, USA}
\author{Xiangdong Ji}
\affiliation{Maryland Center for Fundamental Physics, Department of Physics, \\ University of Maryland, College
Park, Maryland 20742, USA}
\affiliation{Institute of Particle Physics and Cosmology, Department of Physics \\
Shanghai Jiao Tong University, Shanghai, 200240, P.R.China}
\affiliation{Center for High-Energy
Physics and Institute of Theoretical Physics, \\ Peking University
Beijing, 100080, P. R. China}
\author{Yang Xu}
\affiliation{Center for High-Energy
Physics and Institute of Theoretical Physics, \\ Peking University
Beijing, 100080, P. R. China}
\author{Yue Zhang}
\affiliation{Maryland Center for Fundamental Physics, Department of Physics, \\ University of Maryland, College
Park, Maryland 20742, USA}
\affiliation{Center for High-Energy
Physics and Institute of Theoretical Physics, \\ Peking University
Beijing, 100080, P. R. China}

\date{\today}

\begin{abstract}
We study angular momentum in non-relativistic quantum electrodynamics (NRQED). We
construct the effective total angular momentum operator by applying Noether's theorem to
the NRQED lagrangian. We calculate the NRQED matching for the individual components
of the QED angular momentum up to one loop. We illustrate an application of our results
by the first calculation of the angular momentum of the ground state hydrogen atom carried
in radiative photons, $\alpha_{\rm em}^3/18\pi$, which might be measurable in future
atomic experiments.

\end{abstract}

\maketitle

\newcommand{\be}{\begin{equation}}
\newcommand{\ee}{\end{equation}}
\newcommand{\ben}{\[}
\newcommand{\een}{\]}
\newcommand{\beqn}{\begin{eqnarray}}
\newcommand{\eeqn}{\end{eqnarray}}
\newcommand{\Tr}{{\rm Tr} }

Non-relativistic quantum electrodynamics (NRQED) and chromodynamics (NRQCD)
are powerful effective field theories which have been successfully applied
in solving non-relativistic bound-state problems in atomic physics and
QCD heavy quarkonia~\cite{Caswell:1985ui,Brambilla:2004jw}. For example, NRQED has been used
to calculate the hyperfine splitting and lamb shift in QED systems
with considerable ease~\cite{Caswell:1985ui,Kinoshita:1995mt,Luke:1999kz,Manohar:2000rz,Hoang:2001rr}.
NRQCD has been used in analyzing the heavy-quarkonium production in colliders and precision
bound-state calculations in lattice field theory~\cite{Bodwin:1994jh,Lepage:1992tx}.
In this paper, we show how angular momentum can be consistently treated
in these theories, applying the one-loop matching result to calculating the radiative photon
contribution to the spin of the hydrogen atom. The spin
structure of a bound state is a topic of considerable interest in recent years~\cite{Filippone:2001ux}.
The photon contribution to the hydrogen spin is particularly relevant
to probing the gluon contribution to the spin of the proton~\cite{Bunce:2000uv,Chen:2006ng,deFlorian:2008mr}.

The simplest way to obtain the angular momentum operator in NRQCD is to apply
Noether's theorem. The effective NRQED lagrangian to order $1/m^2$ is~\cite{Caswell:1985ui,Manohar:1997qy},
\begin{eqnarray}
  {\cal L}_{\rm eff} &=& \psi^\dagger \left(iD^0 + \frac{{\ma D}^2}{2m}
     +  \frac{{\ma D}^4}{8m^3} + c_F \frac{e}{2m} {\bg \sigma} \cdot {\bf B}
     + c_D \frac{e}{8m^2} (\bf{ D \cdot E - E \cdot D})  \right. \nonumber \\
    &+ & \left. c_S \frac{ie}{8m^2} {\bg \sigma} \cdot (\bf{ D\times E - E\times D}) +...
     \right) \psi
      - \frac{d_1}{4} F_{\mu\nu} F^{\mu\nu}  +
        \frac{d_2}{m^2} F_{\mu\nu} \square
       F^{\mu\nu} + ...
\end{eqnarray}
with $\psi$ the two-component electron field operator and $m$ its mass, $F_{\mu\nu}$ the photon field,
and $D^\mu = \partial^\mu + ie\partial^\mu$, ${\bg \pi} = -i{\bf D}$. To one-loop order, the
matching coefficients are well known, $c_F = 1 + \frac{\alpha_{\rm em}}{2\pi} $, $ c_D = 1 + \frac{8\alpha_{\rm em}}{3\pi}\left( \ln \frac{m}{2\Lambda} + \frac{5}{6}\right)$,
$     c_S = 1 + \frac{\alpha_{\rm em}}{\pi} $, $d_1 = 1-\frac{\alpha_{\rm em}}{3\pi}\ln\frac{\Lambda^2}{m^2}$,
$  d_2 = \frac{\alpha_{\rm em}}{60\pi}$, where $\Lambda$ is the momentum cut-off.
The lagrangian is invariant under three-dimensional rotation which, according to Noether's theorem,
leads to a conserved angular momentum,
\begin{eqnarray}
 \bf{J}_{\rm NRQED}
  &=& \int d^3r  \left[ \psi^\dagger ({\bf r}\times {\bg \pi}) \psi
    + \psi^\dagger \frac{\bg \sigma}{2} \psi + d_1(\bf {r}\times (\bf{E\times B}))\right. \nonumber \\
   &&    + c_D \frac{e}{8m^2} {\bf r}\times ({\bf B} \times \bg{\partial}) (\psi^\dagger \psi)
    - c_S \frac{e}{8m^2} {\bf r} \times ({\bf B}
     \times (\psi^\dagger ({\bg \pi} \times {\bg \sigma})\psi)) + ...
\end{eqnarray}
Apart from the standard non-relativistic contribution, we have $1/m^2$ power corrections as well (we omit the pure
photon contribution at this order for simplicity). These
relativistic effects become important when we analyze the angular momentum structure
of a bound state, as we shall see below.

To understand the spin structure of a bound state, we need to match individual terms
in the full QED angular momentum
\begin{equation}
{\bf J}_{\rm QED}
  = \int d^3r  \left[ \Psi^\dagger ({\bf r}\times \bg{\pi}) \Psi
    + \Psi^\dagger \frac{\bg{\Sigma}}{2} \Psi + (\bf {r}\times (\bf{E\times B}))\right] \nonumber \\
\end{equation}
to the operators in NRQED. To this goal, one has to carry out detailed perturbative matching calculations:
requiring physical amplitudes in both the full and effective theories be the same. For example,
we write down an expansion of the QED spin operator in terms of the operators in NRQED
up to $1/m^2$ order~\cite{chennew},
\begin{eqnarray}
 \left. \Psi^\dagger\frac{\bg \Sigma}{2}\Psi(\mu)\right|_{\rm QED} &=& a_\sigma \psi^\dagger \frac{\bg \sigma}{2}\psi + \frac{a_B}{4m^2} \psi^\dagger e{\bf B}\psi \nonumber
\\  &&  + \frac{a_\pi}{8m^2} \psi^\dagger\left[({\bg\sigma}\times {\bg\pi})\times {\bg\pi} - {\bg\pi}\times({\bg \sigma}\times{\bg \pi})\right]\psi
+...,
\end{eqnarray}
where scale $\mu$ is needed to define the QED operator in dimensional regularization and modified minimal subtraction scheme ($\overline{\rm MS}$),
the infrared-finite constants $a_\sigma$, $a_B$, and $a_\pi$ include
quantum effects at scale $m$ and above and can be calculated as a perturbation series in $\alpha_{\rm em}$.
Likewise, we can write down an expansion for the orbital part,
\begin{eqnarray}
 \left.\Psi^\dagger ({\bf r\times \bg \pi})\Psi(\mu)\right|_{\rm QED} &&=
  d_\sigma\psi^\dag\frac{\bg{\sigma}}{2}\psi + d_R \psi^\dag({\ma r\times \bg{\pi}})\psi + \frac{d_E}{4m}\psi^\dag\left[{\ma r} \times (\bg{\sigma} \times e\ma {E})\right]\psi \nonumber\\
 & & + \frac{d_\pi}{8m^2} \psi^\dag \left[(\bg{\sigma}\times\bg{\pi}) \times \bg{\pi}-\bg{\pi} \times (\bg{\sigma}\times\bg{\pi})\right]\psi + \frac{d_B}{4m^2}\psi^\dag{(e\ma B)}\psi \nonumber\\
 & & + \frac{d_D}{8m^2}  [{\ma r}\times (e\ma B \times \bg{\p})](\psi^\dag\psi) + \frac{d_S}{8m^2}{\ma r}\times\left[(-e\ma{B})\times\big(\psi^\dag (\overleftrightarrow{\bg{\pi}}\times \bg{\sigma}) \psi\big)\right]\nonumber\\
 & & + \frac{d'_S}{8m^2}{\ma r}\times\left[\bg{\sigma}\times\big(\psi^\dag (e\ma {B} \times \overleftrightarrow{\bg{\pi}}) \psi\big)\right] +... \ ,
\end{eqnarray}
and finally the the photon angular momentum operator,
\begin{eqnarray}\label{gamma}
\ma r\times (\ma{E}\times\ma{B})(\mu)|_{\rm QED} & = & d_1(\mu) \ma{r}\times (\ma{E}\times\ma{B})+  f_R \psi^\dag({\ma{r}\times \bg{\pi}})\psi + f_\sigma\psi^\dag\frac{\bg{\sigma}}{2}\psi\nonumber\\
 & & + \frac{f_E}{4m}\psi^\dag\left[{\ma r} \times (\bg{\sigma} \times e\ma {E})\right]\psi + \frac{f_\pi}{8m^2} \psi^\dag \left[(\bg{\sigma}\times\bg{\pi})
\times \bg{\pi}-\bg{\pi} \times (\bg{\sigma}\times\bg{\pi})\right]\psi \nonumber\\
& &  + \frac{f_B}{4m^2}\psi^\dag{(e\ma B)}\psi
+ \frac{f_D}{8m^2}  [{\ma r}\times (e\ma B \times \bg{\p})](\psi^\dag\psi)\nonumber\\
& & + \frac{f_S}{8m^2}{\ma r}\times\left[(-e\ma{B}) \times\big
(\psi^\dag (\overleftrightarrow{\bg{\pi}}\times \bg{\sigma}) \psi\big)\right]\nonumber\\
& & + \frac{f'_S}{8m^2}{\ma
r}\times\left[\bg{\sigma}\times\big(\psi^\dag (e\ma{B} \times
\overleftrightarrow{\bg{\pi}}) \psi\big)\right] + ... \ .
\end{eqnarray}
Note that the non-local angular momentum operators
receive contribution from local operators after matching to non-relativistic theory.
Rotational symmetry imposes the following constraint through the total NRQED angular momentum,
\begin{eqnarray}
    a_\sigma + d_\sigma + f_\sigma &=& d_R + f_R = 1\ ,   \nonumber \\
    a_\pi + d_\pi + f_\pi &= &  a_B + d_B + f_B = 0 \ ,  \nonumber \\
    d_D + f_D &=& c_D  \ , \nonumber \\
    d_S + f_S &=& c_S \ , \nonumber \\
    d_S' + f_S' &=& d_E + f_E = 0 \ .
\end{eqnarray}
Moreover, angular momentum evolution in the full theory imposes constraints
on the renormalization-scale dependence of the coefficients~\cite{Ji:1996ek}.

The various coefficients can be directly evaluated using two- and three-point functions matching.
Through two-point function matching, we get~\cite{chennew},
\begin{eqnarray}
a_\sigma&=&1+\frac{\alpha_{\rm em}}{2\pi};\;\;a_\pi=1+\frac{\alpha_{\rm em}}{2\pi}; \nonumber \\
d_R &=& 1 + \frac{\alpha_{\rm em}}{2\pi}\left(-\frac{4}{3}\ln\frac{\mu^2}{m^2}-\frac{17}{9}\right);\;\; d_\sigma = \frac{\alpha_{\rm em}}{2\pi}\left(-\frac{4}{3}\ln\frac{\mu^2}{m^2}-\frac{20}{9}\right);\;\;d_\pi = -1 - \frac{5\alpha_{\rm em}}{6\pi}\nonumber \\
f_R &=& \frac{\alpha_{\rm em}}{2\pi}\left(\frac{4}{3}\ln\frac{\mu^2}{m^2}+\frac{17}{9}\right);\;\; f_\sigma = \frac{\alpha_{\rm em}}{2\pi}\left(\frac{4}{3}\ln\frac{\mu^2}{m^2}+\frac{11}{9}\right);\;\;f_\pi = \frac{\alpha_{\rm em}}{3\pi}.
\end{eqnarray}
From the three-point function matching, we have~\cite{chennew}
\begin{eqnarray}
a_B&=&1+\frac{7\alpha_{\rm em}}{2\pi}; \nonumber \\
d_D &=& 1 + \frac{\alpha_{\rm em}}{2\pi}\left(-\frac{4}{3}\ln\frac{\mu^2}{m^2} + \frac{16}{3}\ln\frac{m}{2\Lambda} + \frac{37}{9}\right);\;\; d_S = 1+ \frac{\alpha_{\rm em}}{2\pi}\left(-\frac{4}{3}\ln\frac{\mu^2}{m^2}+\frac{31}{9}\right);\nonumber\\
d'_S &=& \frac{2\alpha_{\rm em}}{3\pi};\;\; d_E = \frac{\alpha_{\rm em}}{3\pi};\;\;d_B=-1+\frac{\alpha_{\rm em}}{2\pi} \left(-{16}\ln\frac{m}{2\Lambda}+\frac{23}{9}\right); \nonumber \\
f_D &=& \frac{\alpha_{\rm em}}{2\pi}\left(\frac{4}{3}\ln\frac{\mu^2}{m^2}+\frac{1}{3}\right);\;\; f_S = - \frac{\alpha_{\rm em}}{2\pi}\left(-\frac{4}{3}\ln\frac{\mu^2}{m^2}+\frac{13}{9}\right);\nonumber\\
f'_S &=& -\frac{2\alpha_{\rm em}}{3\pi};\;\;
f_E = -\frac{\alpha_{\rm em}}{3\pi};\;\;f_B=\frac{\alpha_{\rm em}}{2\pi}\left(16\ln\frac{m}{2\Lambda} -\frac{86}{9}\right);
\end{eqnarray}
where $\mu$-dependence comes from the definition of the original QED operators in $\overline{\rm MS}$, and $\Lambda$-dependence comes
from the NRQED calculations with three-momentum cut-off.

As a first application, let us consider the orbital angular momentum
of the electron in the ground state of the hydrogen atom. In non-relativistic theory,
the electron is in $s$-wave with vanishing orbital motion. However, in relativistic framework,
the electron's wave function is a four-component Dirac spinor,
\begin{equation}
  \Psi_{nljm} =\left(\begin{array}{c} \frac{iG_{lj}(r)}{r} \psi_{jm}^l  \\ \frac{F_{lj(r)}}{r}(\sigma\cdot r)\psi_{jm}^l \end{array}\right) \ .
\end{equation}
In the ground state ($n=l=0, j=1/2$), the upper component is an $s$-wave, but the
lower component is a $p$-wave. Therefore, there is an orbital angular momentum contribution to the ground state
spin of the hydrogen atom (we ignore the spin of the proton) which can be calculated directly through the above
Dirac wave function. In effective NRQED, this angular momentum contribution can be calculated as the matrix elements of the
\begin{equation}
           -\frac{d_\pi}{8m^2}\int d^3r \psi^\dagger  [({\bg\sigma} \times {\bg \pi})\times {\bg \pi}
             - {\bg \pi} \times ({\bg \sigma}\times {\bg \pi})]\psi \ ,
\end{equation}
in the Coulomb wave function, which yields easily $ \langle L_z \rangle = \alpha^2_{\rm em}/6 $,
consistent with the full theory. This contribution is balanced by the equal amount of
depletion of the electron spin contribution $\langle \Sigma_z/2\rangle = 1/2-\alpha^2_{\rm em}/6$.

\begin{figure}[ht]
\begin{center}
\includegraphics[height=3cm]{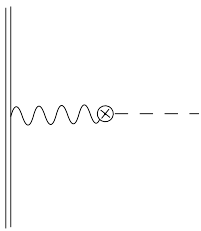}
\end{center}
\vskip -0.7cm \caption{Leading (${\cal O}$($\alpha^2_{\rm em}$)) electromagnetic contribution to the spin of the Coulomb-bound electron.
The cross represents the photon angular momentum operator; the dashed line an external Coulomb field and the wavy
line an off-shell photon field. The double line is an electron eigenstate in the external static Coulomb field.}
\end{figure}

A more interesting question is: what is the amount of hydrogen spin carried by radiative photons? This question
is particularly relevant in the spin structure of the nucleon \cite{Ji:1996ek,Filippone:2001ux} since, due to strong coupling, the QCD gluon could contribute significantly to the nucleon
spin~\cite{Chen:2006ng,Jaffe:1995an}. The spin program at the Relativistic Heavy-Ion Collider is largely
motivated by this possibility~\cite{Bunce:2000uv}. Here to simplify the problem,  we take the proton mass to
infinity and are left with essentially an electron in a static Coulomb potential.

To answer this question, let us first consider the calculation in the full QED theory.
The leading order contribution comes from the diagram in Fig. 1.
This contribution can be easily shown to be zero because of the Coulomb nature
of the static potential~\cite{manohar}, ${\bf E} = {\bg {\nabla}} e/(4\pi r)$,
\begin{equation}
 \int d^3r ~{\bf r\times (E\times B)} = 0  \ .
\end{equation}
Therefore, the magnetic field generated by the electron
current does not contribute to the hydrogen's angular momentum. This contribution,
were it non-zero, would have been of order $\alpha^2_{\rm em}$.

Thus, the first non-vanishing electromagnetic contribution comes
from radiative photons shown by Fig. 2a, which will be at least order
$\alpha^3_{\rm em}$.  The loop integral is ultraviolet divergent by simple
power counting. One can interpret this divergence in two ways: First, define
photon and electron in terms of a certain renormalization (or cut-off) scheme,
and the answer is finite within the scheme. But the result is then scheme
and scale dependent~\cite{Ji:1996ek}. This, however, is the preferred approach in QCD because
there are no free quarks and gluons due to color confinement.
A second approach is to define the electron and photon using the asymptotic physical states.
In this case, the physical electron spin acquires the same radiative corrections and therefore
one must subtract off the contribution in which the intermediate
electron is a free-space one, as shown by Fig. 2b. The subtraction will produce a
finite contribution, i.e., free of ultraviolet divergence. This situation is similar to the famous Lamb shift calculation for the energy shift.

\begin{figure}[ht]
\begin{center}
\includegraphics[height=4.3cm]{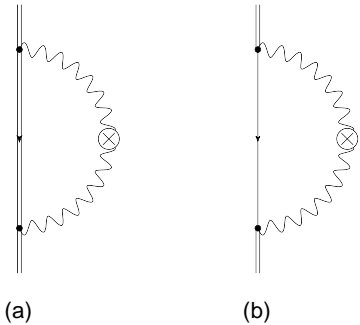}
\end{center}
\vskip -0.7cm \caption{a) Next-to-leading order (${\cal O}$($\alpha^3_{\rm em}$))  electromagnetic contribution to the spin of the
Coulomb-bound electron, b) subtraction needed to define the physical contribution.}
\end{figure}

To calculate the contribution from Fig. 2a, we use the NRQED approach similar to the calculation
of the Lamb shift outlined in Ref. \cite{holstein}.
We split the loop momentum in Fig. 2a into small and large regions.
When the loop-momentum is large, we can expand the bound state
electron wave function in terms of its successive interactions with
the static Coulomb field. After subtracting off the free contribution, we are left
with Fig. 3 in the large loop momentum region.
This contribution can be matched to local operators made of the quark fields in Eq. (5),
related to the matching coefficients $f_E$. Therefore, the large
momentum contribution is
\begin{equation}
\label{total}
  \left\langle \int d^3r {\bf r}\times ({\bf E}\times {\bf B}) \right\rangle
   = \left \langle \frac{f_E}{4m} \int d^3r \psi^\dagger[{\bf r} \times ({\bg\sigma}\times e{\bf E})]\psi\right\rangle
   = \frac{\alpha^3_{\rm em}}{18\pi} \ .
\end{equation}
There is no logarithm associated with this, in contrast to the Lamb shift.

\begin{figure}[ht]
\begin{center}
\includegraphics[height=3.8cm]{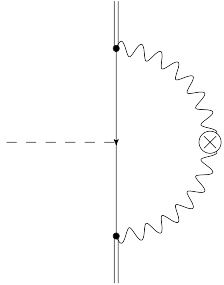}
\end{center}
\vskip -0.7cm \caption{First-order expansion of electromagnetic contribution to the spin in an external Coulomb field, in large loop momentum region.}
\end{figure}

In the low-momentum region, we calculate the matrix elements of $\int d^3r {\bf r} \times ({\bf E} \times {\bf B})$
using the old-fashioned first-order perturbation theory. It is easy to see that the
contribution is zero, including the free intermediate state contribution. This
is because Figs. 2a and 2b in NRQED are independent of the electron spin.
Therefore, Eq. (\ref{total}) is the total radiative photon contribution to the spin of the Coulomb
bound electron. It is interesting to note
that the result is positive. Future atomic physics experiments
might be able to measure this small quantity. To maintain the total spin 1/2, this
photon contribution is balanced by the electron's orbital motion, as is clear from Eq. (4).

In conclusion, we have calculated the matching of the angular momentum
operator to that in non-relativistic effective theory. Using the tree-level result, we easily
reproduce the orbital angular momentum in the s-wave state using Coulomb
wave functions. Using the one-loop matching for the photon angular
momentum, we calculated the amount of the hydrogen angular
momentum carried in the radiative photons, which is a positive $\alpha^3_{\rm em}/18\pi$.

We thank D. Beck for initiating our interest in this project and D. Beck, B. Holstein and A. Manohar for useful discussions.
This work is partly supported by the U. S. Department of Energy via grant DE-FG02-93ER-40762.

\end{document}